\documentclass[prl,aps,twocolumn]{revtex4-1}  % DON'T CHANGE
\begin{document}                % INITIALIZE - DONT CHANGE
\title{A dynamical equation for a maser
with non-poissonian injection statistics}
\author{Michael Fleischhauer}
\address{Dept. of Physics and research center OPTIMAS \\
University of Kaiserslautern, 67659 Kaiserslautern, Germany
}
% address. (Remove the left % marks)
%

%\date{\today}

\begin{abstract}                % DON'T CHANGE THIS LINE
A derivation of the coarse grained dynamical equation  for  
a maser with periodic injection of atoms 
suggested by Briegel and Englert [Phys.Rev.A. {\bf 52}, 2361 (1995)] 
from the microscopic masterequation is presented. 
\end{abstract}
\maketitle
%\pacs{42.50-p, 42.55.-f,32.80.-t,05.30-t}
%

The description of the dynamics of a maser with non-poissonian 
injection statistics of atoms 
has been the subject of several discussions in the
literature \cite{maserpump}. While a microscopic description of 
the field evolution by a sequence of kicks from the
atoms and subsequent periods of free decay adeqately models 
the maser dynamics \cite{micro}, several approaches to derive a 
quasi-continous masterequation on a coarse grained time-scale failed 
for non-poissonian pumping \cite{maserprobl}. Briegel and Englert 
suggested a macroscopic masterequation which overcomes these 
problems \cite{BrEng}. 
\begin{equation}
\frac{\partial}{\partial t} \bar\rho(t) = {\cal L} \bar\rho(t) + {\cal K} \frac{\cal L}{1-e^{-{\cal L}T}}\bar\rho(t) .
\end{equation}
The aim of this short note is to derive this
equation from the microscopic masterequation. 

We here consider the following microscopic model of the maser: 
Some pumping mechanism periodically injects excited two-level atoms into a
resonator. 
The number of atoms entering the cavity at the periodic times $t_j=jT$ 
$(j=0,\pm 1,\dots)$ can fluctuate. 
The probability for a $k$-atom event  
is denoted by $p_k$. If we assume, that the transit time of 
the atom(s) is short compared 
to the time scale of interest, the effect of the atoms on the field
can be described by a sequence of quasi instantaneous kicks
\begin{equation}
\rho(jT+0)=(1+{\cal K})\rho(jT-0)\label{kick}
\end{equation}
where
\begin{equation}
{\cal K}=\sum_{k=1}^\infty p_k {\cal M}_k
\end{equation}
is the operator describing the average effect of the injected 
atoms on the field. 
${\cal M}_k$ accounts for the change in the field resulting from a
$k$-atom event and  is not further specified here. It depends on the 
actual interaction process and the passage time.
Denoting the Liouvillian that describes the coupling 
to the cavity reservoir by ${\cal L}$ the microscopic masterequation for 
the field evolution in the interaction picture reads:
\begin{equation}
\frac{\partial}{\partial t} \rho(t)={\cal L}\rho(t)+\lim_{\epsilon\to
+0}{\cal K}\sum_j\delta(t-jT) 
\rho(t-\epsilon).\label{mikro}
\end{equation}

In order to derive an equation of motion on a coarse grained time scale we 
introduce a time averaged density operator:
\begin{equation}
{\bar\rho}(t)=\int_{-T_0/2}^{T_0/2} d\tau\rho(t-\tau)f(\tau),\label{rhoeff}
\end{equation}
where $f(\tau)$ is a properly normalized, slowly varying function of time.
The averaging interval $T_0$ is assumed to be larger, or of the order of,
the injection period $T$. 
It is clear at hand, that we can not define a coarse grained
density operator in the immediate vicinity of the initial time $t=0$.
This is a generic feature of any coarse-graining approximation to an
initial value problem and the definition of ${\bar \rho}(t)$ makes only sense for times larger
than the averaging interval $T_0/2$.

In this sense we may rewrite Eq.(\ref{rhoeff}) in a form
convenient for a Laplace-transformation
\begin{equation}
{\bar\rho}(t)=\int_0^{t} d\tau\rho(t-\tau)f(\tau).\label{rhoeff2}
\end{equation}
We now use this equation as a definition of a coarse-grained density
operator, noting that it has the correct properties of a density operator
only for time $t\ge T_0$. 
The Laplace-transform of $\bar\rho$ 
is then simply obtained from that of $\rho$ via 
$ {\bar\rho}(s)=\rho(s)f(s)$, 
where, for notational simplicity, 
we used the same symbols for the functions in Laplace-space. 

We proceed by transforming the microscopic masterequation (\ref{mikro}).
The multiplication with the ``filterfunction'' $f(s)$ will allow some
approximations which eventually yield the desired macroscopic masterequation.
\begin{eqnarray}
(s-{\cal L})&&{\rho}(s)=\rho(t=0)\label{rho1} \\
+\lim_{\epsilon\to +0}&&\int_{0}^\infty
dt \sum_j \delta (t-jT) e^{-st}{\cal K}\rho(t-\epsilon).\nonumber
\end{eqnarray}
The sum of delta-functions in Eq.(\ref{rho1}) is equivalent to a sum of
exponentials
\begin{equation}
\sum_{j=-\infty}^\infty \delta(t-jT)=\frac{1}{T}
\sum_{\nu=-\infty}^\infty e^{2\pi i\nu t/T}
\end{equation}
which yields
\begin{eqnarray}
(s-{\cal L})&&{\rho}(s)=\rho(t=0)\label{rho2}\\ 
&&+\lim_{\epsilon\to+0}\frac{{\cal K}}{T}\sum_\nu
\int_0^\infty dt\enspace e^{-(s-2\pi i\nu/T)t}
\rho(t-\epsilon).\nonumber
\end{eqnarray}
In Eq.(\ref{rho2}) we can immediately identify the Laplace-transform of $\rho$
with a shifted argument:
\begin{eqnarray}
(s-{\cal L})&&{\rho}(s)=\rho(t=0)\label{rho3}\\
&&+\lim_{\epsilon\to+0}\frac{{\cal K}}{T}\sum_\nu
e^{-(s-2\pi i\nu/T)\epsilon}
\rho(s-\frac{2\pi i\nu}{T}).\nonumber
\end{eqnarray}
Since the r.h.s.~of Eq.(\ref{rho3}) is invariant under the transformation
$s\to s+2\pi i n/T$ with $n=0,\pm 1,\dots$, the l.h.s~is invariant as
well.
From this we infer
\begin{equation}
{\rho}(s+\frac{2\pi i\nu}{T})=\left[s+\frac{2\pi i\nu}{T}-{\cal
L}\right]^{-1}(s-{\cal L})\ {\rho}(s).\label{rho4}
\end{equation}
Inserting this result into Eq.(\ref{rho3}) yields 

\begin{widetext}
\begin{eqnarray}
(s-{\cal L})\rho(s)-\rho(t=0)
&=&\lim_{\epsilon\to+0}\frac{{\cal K}}{T}\sum_{\nu=-\infty}^\infty
\frac{s-{\cal L}}{s+\frac{2\pi i\nu}{T}-{\cal L}}\
{\rho}(s)
\ e^{-(s+2\pi i \nu/T)\epsilon}\label{rho5}\\
&=&\lim_{\epsilon\to+0}\frac{{\cal K}}{T}\sum_{\nu=-\infty}^\infty
\left[ 1-\frac{i\frac{2\pi\nu}{T}}{s+\frac{2\pi i\nu}{T}-{\cal
L}}
\right]\ {\rho}(s)
\ e^{-(s+2\pi i \nu/T)\epsilon}.\nonumber
\end{eqnarray}

\end{widetext}

So far no approximations are made. 
We now multiply Eq.(\ref{rho5}) with the filterfunction $f(s)$, which
rapidely decreases for $|s|>T_0^{-1}$. If we take an averaging intervall 
long compared to the period of injection $T$, we may therefore neglect $s$
in the denominator of Eq.(\ref{rho5}) as compared to $2\pi i\nu/T$.
Note that the second term in the brackets vanishes for $\nu=0$.

If we furthermore rewrite the sum over $\nu$
as a sum of integrals with the help of the Poisson summation formula
\cite{Poisson},
we obtain:
\begin{eqnarray}
(s &&-{\cal L}){\bar\rho}(s)=\rho(t=0)f(s)+\lim_{\epsilon\to
+0}\frac{{\cal K}}{T}\times\label{rho6}\\
&&\times\int_{-\infty}^\infty d\nu \sum_{l=-\infty}^\infty
e^{-i2\pi\nu(\epsilon/T+l)}
\left[1-\frac{\nu}{\nu+i\frac{{\cal L} T}{2\pi}}\right]
{\bar\rho}(s).\nonumber
\end{eqnarray}
The $\nu$ integration of the first term gives 
deltafunctions $\delta(\epsilon/T+l)$ which vanish for $l=0,\pm 1,\dots$. The
$\nu$ integration of the second term can be carried out by residual
integration.
Noting, that the eigenvalues of ${\cal L}$ are zero or negative, 
we find that only
terms with negative $l$ values contribute. 
We thus have
\begin{eqnarray}
(s-{\cal L}){\bar\rho}(s)-\rho(t=0)f(s)&=&-{\cal K}
\sum_{l=-\infty}^{-1} {\cal L} e^{-{\cal L} T l}
{\bar\rho}(s)\nonumber\\
&=&-{\cal K}\sum_{l=1}^\infty {\cal L} e^{{\cal L} T l}
{\bar\rho}(s)\label{rho7}\\
&=&{\cal K}\frac{{\cal L}}{1-e^{-{\cal L} T}}\
{\bar\rho}(s).\nonumber 
\end{eqnarray}
A transformation into the time domain yields 
\begin{eqnarray}
\frac{\partial}{\partial t}{\bar\rho}(t)&=&\rho(t=0)f(t)-{\bar\rho}(t=0)\\
&&+{\cal L}{\bar\rho}(t)
+{\cal K}\frac{{\cal L}}{1-e^{-{\cal L} T}}\
{\bar\rho}(s).\nonumber 
\end{eqnarray}
Noting, that according to Eq.(\ref{rhoeff2}) ${\bar\rho}(t=0)=0$ and that for
$t>T_0$ $f(t)\equiv 0$, we obtain the macroscopic masterequation
of Briegel and Englert \cite{BrEng}:
\begin{equation}
\frac{\partial}{\partial t}{\bar\rho}(t)={\cal L}{\bar\rho}(t) 
+{\cal K}\frac{{\cal L}}{1-e^{-{\cal L}T}}\ {\bar\rho}(t).
\end{equation}
Since Eq.(\ref{rhoeff2}) defines a correct coarse-grained density operator
only for $t\ge T_0$, this equation is only true for these times. In order to
find the correct initial value of ${\bar\rho}(t)$ one has to solve the 
microscopic equation for some small time intervall and calculate ${\bar\rho}(T_0)$
according to Eq.(\ref{rhoeff}).
If the dynamics is however sufficiently slow one may to a good approximation
apply the macroscopic equation also in the initial time period and identify the
macroscopic initial value with the microscopic one.

\

\end{document}